\documentclass[aps,preprintnumbers,onecolumn,amsmath,amssymb,floatfix,pra]{revtex4} 
\pdfoutput=1
\usepackage[utf8]{inputenc}
\usepackage{amsmath}
\usepackage{amsfonts}
\usepackage{braket}
\usepackage{tensor}
\usepackage{graphicx}
\usepackage[colorlinks=true,linkcolor=blue, citecolor=blue, bookmarks]{hyperref}
\def\be{\begin{equation}}
\def\ee{\end{equation}}
\def\bea{\begin{eqnarray}}
\def\eea{\end{eqnarray}}
\def\fr#1{(\ref{#1})}

\begin{document}

\title{Quantum quenches in the sinh-Gordon model: steady state and one point correlation functions}
\author{Bruno Bertini}
\author{Lorenzo Piroli}
\author{Pasquale Calabrese}
\affiliation{SISSA and INFN, via Bonomea 265, 34136 Trieste, Italy. }

\begin{abstract}
We consider quantum quenches to the sinh-Gordon integrable quantum field theory from a particular class of initial states. Our analysis includes the case of mass and interaction quenches starting from a non-interacting theory. By means of the recently developed quench action method, we fully characterize the stationary state reached at long times after the quench in terms of the corresponding rapidity distribution. We also provide exact results for the expectation values of arbitrary vertex operators in the post-quench stationary state by proposing a formula based on the analogy with the standard thermodynamic Bethe ansatz. Finally, we comment on the behavior of the post-quench stationary state under the mapping between the sinh-Gordon field theory and the one-dimensional Lieb-Liniger model.
\end{abstract}
\maketitle

\section{Introduction}\label{sec:intro}

There exists an intimate connection between integrable quantum field theories and one-dimensional integrable models in many-body quantum physics. Indeed, it is hard to disentangle their history and developments over the last fifty years \cite{thacker, faddeev, korepin, ek-review}. During the past decade, the advent of ultra-cold atoms \cite{bloch-review, bdz-08, ccgor-11} has opened the way to the experimental realization of nearly ideal one-dimensional integrable quantum systems, triggering new efforts in understanding the physical implications of integrability. As a natural consequence, integrable quantum field theories have then recently enjoyed a refreshed theoretical interest. This is particularly true in connection with the study of the non-equilibrium dynamics of isolated quantum systems, which is the object of an intense on-going research activity \cite{pssv-11, GE15}. 

From the theoretical point of view, a relevant body of literature has been focusing on the determination of the stationary value reached by local correlations of the systems at late times after a quantum quench, i.e. a sudden change in the Hamiltonian parameters. In this context the behavior of integrable systems is special because of the presence of an infinite set of local conserved operators (or charges) constraining the dynamics at all times. It is then natural to conjecture that the asymptotic properties of the system are captured by a generalized Gibbs ensemble (GGE) taking into account all the conserved charges \cite{rigol-07}. The GGE construction allows to predict the physical properties of a macroscopic system without explicitly solving the whole non-equilibrium dynamics. Indeed, one takes into account only a limited amount of information on the initial state (encoded in the expectation value of the conserved charges) to characterize the eventual stationary value of correlations of the system.

In the past years, an impressive amount of work has been devoted to test the GGE and understand its range of applicability \cite{cazalilla-06, cc-07, barthel-08,eckstein, cdeo-08, rigol-09, iucci-09,fioretto-10,cramer,cassidy,foini,cef-11,eef-12,ck-12,mossel,gramsch,
dora,fagotti-13,pozsgay-13_II,goldstein-13,gurarie,kormos-13,collura,
mussardo,fagotti-14,fcec-14,sotiriadis,mazza,pozsgay,bucciantini,goldstein, bf-14,alba-15, bastianello}. In order to obtain exact predictions on the asymptotic state, the GGE approach requires the {\it a priori} knowledge of a complete set of local or quasilocal conserved charges of the model at hand (at least for the sector of the Hilbert space specified by the symmetries of the initial state). In general this is a highly non-trivial problem, as demonstrated by the recent discoveries of additional
conservation laws in integrable quantum spin chains \cite{fagotti-14, prosen-11, pi-13, ppsa-14, imp-15, idwc-15, iqdb-15, pv-16, fagotti-16} and integrable field theories \cite{emp-15, cardy-15}.

A different approach has recently been proposed to describe the long-times steady state following a quench in integrable systems: the so called quench action method (QAM), a.k.a. representative eigenstate approach \cite{ce-13}. The latter has proven to be a powerful and versatile technique which has been applied in the study of quenches in Heisenberg spin chains \cite{wdbf-14, pmwk-14,ac-15}, interacting Bose gases \cite{dwbc-14, pce-15, bucciantini-15}, quantum field theories \cite{bertini} and transport problems \cite{dmv-15}. In the recent works \cite{bertini, dc-14,pdc-15, vwed-15} it has been possible to apply in some special cases the QAM to tackle the computation of the whole time evolution of local observables in interacting models. However, this remains in general an extremely difficult task. Explicit results for the computation of the whole post-quench time evolution were also obtained by independent approaches in Refs.~\cite{collura, cazalilla-06, cc-07, iucci-09,rsm-12, se-12, ia-12, goldstein-13, delfino-14,kcc-14,kz-15}.

As we will see in the next section, the difficulties in the application of the QAM are almost entirely encoded in the exact determination of the initial state in the basis of the post-quench Hamiltonian. No general scheme has been developed yet to tackle this problem which for the moment remains to be analysed case by case \cite{kozlowski, pozsgay-14, bdwc-14, pc-14, mazza-15, fz-15, mpc-10, brockmann-14, sfm-12, stm-14, hst-16}.

Within the QAM, the post-quench stationary state is identified with an excited representative eigenstate of the post-quench Hamiltonian, rather than a statistical ensemble \cite{ce-13}. The asymptotic stationary value of the time-dependent local correlation functions can then be obtained as the expectation value of the corresponding local operators on this representative eigenstate. Note, however, that the computation of these correlation functions remains generically an open problem. This scenario is common in many-body physics: even in integrable quantum field theories, where the scattering matrix is exactly known, the computation of correlation functions is an extremely difficult task, which has been investigated mainly for the vacuum state or at finite temperature \cite{zamolodchikov-91, dm-94, dms-95, llss-96, dsc-96, lz-97, lm-99, saleur-00, lukyanov-01, delfino-01, mussardo-01, c-af-02, delfino-04, doyon-05, pt-08, takacs-08, ek-09, kp-10, ns-13, negro-14}. In the study of quantum quenches, the physical characterization of the eventual stationary state requires the computation of correlation functions in generic, non-thermal, highly excited states. This problem has not been addressed until very recently and only in the past few years some progress has been made in this direction for the simplest case of one-point correlation functions \cite{pozsgay-11, pozsgay2-11, pozsgay-13, mp-14, ga-15, pst-15}.

In Ref.~\cite{pozsgay-11} it was shown that in massive integrable quantum field theories with diagonal scattering matrix, one-point functions for excited states can be calculated using the LeClair-Mussardo formalism \cite{lm-99}. This is a non-trivial achievement, since the latter was originally conceived and derived to compute expectation values at finite temperature \cite{lm-99}. Importantly, the application of the LeClair-Mussardo series has recently led to remarkable results for one-point functions in the one-dimensional $\delta$-interacting Bose gas (namely, the Lieb-Liniger model), both for thermal and arbitrary excited states \cite{kmt-09, kmt-11,kci-11}. These results are based on a mapping between the sinh-Gordon field theory and the Lieb-Liniger model, which is the non-relativistic limit of the former \cite{kmt-09}.

Despite its conceptual significance, the application of LeClair-Mussardo formalism in relativistic field theories is in general still not satisfactory from a computational point of view. Indeed, far from perturbative regimes, explicit numerical evaluation of the LeClair-Mussardo series requires to retain a large number of terms, each term involving the computation of non-trivial multiple integrals. In the sinh-Gordon model, an alternative approach to obtain thermal expectation values of arbitrary vertex operators was recently proposed in Refs.~\cite{ns-13, negro-14}, and explicit expressions for the latter were derived. The advantage of these formulae lies in the fact they do not involve infinite series and can be numerically evaluated for arbitrary values of the temperature with high efficiency.

In this work we focus on the sinh-Gordon model and we address different issues discussed in this introduction. In particular, we consider quantum quenches from a particular class of initial states, which include the case of mass and interaction quenches from the non-interacting theory. By means of the QAM, we fully characterize the stationary state reached at late times and we explicitly verify that the generalized Gibbs ensemble constructed using the conserved rapidity occupation numbers provides the same predictions of the QAM for the long-times behavior of the system. Motivated by the work of Refs.~\cite{pozsgay-11}, and based on the finite temperature results of Refs.~\cite{ns-13, negro-14}, we propose a formula for one-point correlation functions for arbitrary excited states in the sinh-Gordon model. It is shown to provide the correct result in some known cases and it is applied to the quench problem investigated in this work. Finally, we discuss the behavior  of the eventual post-quench stationary state under the mapping to the Lieb-Liniger model. 

The manuscript is organized as follows. In section \ref{sec:shG} we introduce the sinh-Gordon model and discuss the initial state of interest in this work. The QAM is then reviewed in section \ref{sec:steady_state} where we derive and solve the saddle-point equations describing the post-quench stationary state. In section \ref{sec:one_point} we propose a formula for the expectation value of arbitrary vertex operators on excited states and apply it to compute one point correlation functions in the long-times steady state. Finally, in section \ref{sec:scaling} we discuss the mapping to the the Lieb-Liniger model of a mass quench in the sinh-Gordon theory. Conclusions are presented in section \ref{sec:conclusions}.


\section{The sinh-Gordon model and the initial states}\label{sec:shG}

\subsection{The model and the Bethe equations}\label{model}
The sinh-Gordon model corresponds to a relativistic integrable field theory in $1+1$ dimensions. It is described by the Hamiltonian
\be
H=\int{\rm d}x\left\{\frac{c^2}{2}\pi^2(x)+\frac{1}{2}\left[\partial_x \phi(x)\right]^2+\frac{\mu^2c^2}{g^2}:\!\cosh\left[ g\phi(x) \right]\!:\right\},
\label{hamiltonian}
\ee
where $\phi(x)$ is a real scalar quantum field and $\pi(x)$ its conjugate momentum satisfying
\be
\left[\phi(x),\pi(y)\right]=i\delta(x-y).
\ee
The constant $c$ in \fr{hamiltonian} denotes the speed of light and $:\!\ldots\!:$ denotes the normal ordering with respect to the ground state. This integrable field theory is characterized by one single particle species with physical renormalized mass $m$. It is related to the parameter $\mu$ in \fr{hamiltonian} by \cite{kmt-11, bk-02}
\be
m^2=\mu^2\frac{\sin(\alpha\pi)}{\alpha\pi}\,,
\label{Eq:alphamass}
\ee
where
\be
\label{Eq:alpha}
\alpha=\frac{cg^2}{8\pi+cg^2},
\ee
is the renormalized coupling constant. The Hamiltonian \fr{hamiltonian} is left invariant by the following $\mathbb{Z}_2$ transformation
\be
\label{Eq:Z2}
T:\phi(x)\rightarrow-\phi(x)\,.
\ee

In the infinite system the (massive) excitations above the vacuum state $|\Omega\rangle$ can be described in terms of the Zamolodchikov-Faddeev operators, denoted by $Z^{\dagger}(\theta)$, $Z(\theta)$. They obey the so called Zamolodchikov-Faddeev algebra, which in our case reads
\bea
Z(\theta_1)Z(\theta_2)&=&S(\theta_1-\theta_2)Z(\theta_2)Z(\theta_1) \label{zf_I},\\
Z^{\dagger}(\theta_1)Z^{\dagger}(\theta_2)&=&S(\theta_1-\theta_2)Z^{\dagger}(\theta_2)Z^{\dagger}(\theta_1) \label{zf_II},\\
Z(\theta_1)Z^{\dagger}(\theta_2)&=&S(\theta_2-\theta_1)Z^{\dagger}(\theta_2)Z(\theta_1)+2\pi\delta(\theta_1-\theta_2),\label{zf_III}
\eea
where $S(\theta)$ is the S-matrix of the sinh-Gordon model
\be
S(\theta)=\frac{\sinh\theta-i\sin(\alpha\pi)}{\sinh\theta+i\sin(\alpha\pi)}.
\label{s-matrix}
\ee 

A powerful method to analyse two dimensional integrable quantum field theories is provided by the thermodynamic Bethe ansatz (TBA) \cite{zamolodchikov-90,zamolodchikov-00}, which we now briefly review within the sinh-Gordon model. The starting point is the quantization of the theory on a finite ring of length $L$. Since the number $N$ of quasi-particles (or massive excitations) is well defined, one can restrict to the sector corresponding to a specific value of $N$. The quasi-particles are then characterized by the rapidities $\{\theta_j\}_{j=1}^{N}$. Imposing periodic boundary conditions, the quantization procedure leads to the Bethe equations
\be
m c\sinh\theta_j=\frac{2\pi I_j}{L}-\frac{1}{L}\sum_{\substack{k=1\\ k\neq j}}^{N} \phi\left(\theta_j-\theta_k\right),\qquad j=1,\ldots N,
\label{bethe_eq}
\ee
where
\be
\phi(\theta_j-\theta_k)\equiv-i\ln S(\theta_j-\theta_k),
\ee
with $S(\theta)$ given by \eqref{s-matrix}. Here, $\{I_j\}_{j=1}^N$ are a set of integer or semi-integer numbers that can be viewed as a complete set of quantum numbers specifying the global state of the system. For further reference we also introduce the following functions
\bea
Q_j(\theta_1,\ldots,\theta_N)&=&Lmc\sinh\theta_j+\sum_{\substack{k=1\\ k\neq j}}^{N} \phi\left(\theta_j-\theta_k\right),\qquad j=1,\ldots N,\label{q_function}\\
\overline{Q}_j(\theta_1,\ldots,\theta_N)&=&Lmc\sinh\theta_j+\sum_{\substack{k=1\\ k\neq j}}^{N} \phi\left(\theta_j-\theta_k\right)+\sum_{k=1}^{N} \phi\left(\theta_j+\theta_k\right),\qquad j=1,\ldots N.\label{q_bar_function}
\eea
The TBA construction then proceeds to consider the Bethe equations \eqref{bethe_eq} in the thermodynamic limit defined by $L, N\to\infty$, where $D=L/N$ is kept constant. In order to do so, one introduces the quasi-particle rapidity distribution $\rho(\theta)$, together with the hole rapidity distribution $\rho^{h}(\theta)$. The latter is the distribution of unoccupied states and it is analogous to the well known distribution of holes in the Fermi gas at finite temperature. Following the standard derivation \cite{korepin}, Eq. \fr{bethe_eq} leads to 
\be
\rho(\theta)+\rho^h(\theta)=\frac{mc}{2\pi}\cosh(\theta)+\int_{-\infty}^{\infty}\frac{{\rm d}\theta'}{2\pi}\rho(\theta')\varphi_{\alpha}(\theta-\theta'),
\label{thermo_bethe_eq}
\ee
where
\be
\varphi_{\alpha}(\theta)\equiv\frac{{\rm d}}{{\rm d}\theta}\left(-i\ln S(\theta)\right)=\frac{2\cosh(\theta)\sin(\alpha\pi)}{\sinh^2(\theta)+\sin^2(\alpha\pi)}.
\label{Eq:varphi}
\ee
In the following we also define
\be
\eta(\theta)=\frac{\rho^h(\theta)}{\rho(\theta)}.
\label{def_eta}
\ee
Note that the sinh-Gordon model does not exhibit particle bound states. Accordingly, we have only one quasi-particle rapidity distribution $\rho(\theta)$ within the TBA description.

\subsection{The initial states}\label{sec:initial}

In this section we discuss the class of the initial states considered in this work. As we briefly mentioned in the introduction, one of the main technical challenges in the application of the QAM method consists in the specification of the initial state in the basis of the post-quench Hamiltonian. This problem has been addressed mainly in non-relativistic models such as Heisenberg spin chains \cite{kozlowski, pozsgay-14, bdwc-14, pc-14, mazza-15, fz-15} and delta-interacting Bose gases \cite{mpc-10, brockmann-14, dwbc-14, bucciantini-15} and it is even more challenging in the case of integrable quantum field theories \cite{fioretto-10, sfm-12,stm-14,hst-16}. Indeed, the ground state (or vacuum state) of the pre-quench Hamiltonian typically does not have a well defined number of post-quench excitations and the relation between the Zamolodchikov-Faddeev operators corresponding to the theories before and after the quench in general is not known \cite{sfm-12}.

In this work we consider initial states of the form
\be
|\Psi_0\rangle =\exp\left(\int_0^{\infty}\frac{{\rm d}\theta}{2\pi}K(\theta)Z^{\dagger}(-\theta)Z^{\dagger}(\theta)\right)|\Omega\rangle,
\label{initial}
\ee
where $K(\theta)$ is an arbitrary function and $Z^{\dagger}(\theta)$ are the Zamolodchikov-Faddeev operators satisfying Eq. \fr{zf_II}. The class of states \fr{initial}, which are sometimes called squeezed coherent states, have been investigated in a number of recent works \cite{fioretto-10, sfm-12, stm-14, hst-16, bertini, kz-15} and include the integrable boundary states studied in Ref. \cite{gz-94}. Our interest in the class of states \fr{initial} is two-fold. On the one hand, their form is sufficiently simple, although still rather general, to allow for analytic calculations within the framework of integrable quantum field theory \cite{fioretto-10, bertini, se-12, kz-15}. On the other hand, in Refs.~\cite{stm-14, hst-16} evidence has been presented that the form of the initial states for the physically interesting situation of mass and interaction quenches in the sinh-Gordon field theory is given, at least approximately, by Eq. \fr{initial}.

More precisely, consider a free pre-quench quantum field theory, characterized by the Hamiltonian
\be
H=\int{\rm d}x\left\{\frac{c^2}{2}\pi^2(x)+\frac{1}{2}\left[\partial_x \phi(x)\right]^2+\frac{m^2_0c^2}{2}\phi(x)^2\right\}.
\label{hamiltonian_free}
\ee
Suppose that the post-quench Hamiltonian is given by Eq. \fr{hamiltonian}, with $m\neq m_0$ and a non vanishing coupling $\alpha\neq 0$ [the constant $\alpha$ is defined in Eq. \fr{Eq:alpha}]. Then, it was proposed in Ref.~\cite{stm-14} that the initial state of this quench is of the form \fr{initial}, where the function $K(\theta)$ is given by
\be
K(\theta)=K_{D}(\theta)\left(\frac{E_0(\theta)-E(\theta)}{E_0(\theta)+E(\theta)}\right)\,.
\label{k_function}
\ee
Here
\be
K_{D}(\theta)=i\tanh(\theta/2)\left(\frac{1+\cot(\pi\alpha/4-i\theta/2)}{1-\tan(\pi\alpha/4+i\theta/2)}\right),
\ee
and
\be
E(\theta)=\sqrt{m^2c^4\sinh^2\theta+m^2c^4},\qquad E_0(\theta)=\sqrt{m^2c^4\sinh^2\theta+m_0^2c^4}.
\ee
Note that if the initial and final physical masses were equal, namely $m=m_0$, Eqs. \fr{initial}, \fr{k_function} would imply that the vacuum states of the pre-quench and post-quench theories coincide and no dynamics take place. In the limit of an infinite initial mass $m_0\to\infty$ we have $K(\theta)=K_{D}(\theta)$ and the state in Eq. \fr{initial} coincides with the Dirichlet integrable boundary state \cite{gz-94,stm-14}. In Ref.~\cite{stm-14} the ansatz \fr{k_function} was proposed as the initial state of the mass and interaction quench described above in the limit $m_0\gg m$. More recently, in Ref.~\cite{hst-16} systematic evidence in favour of the ansatz \fr{k_function} was provided,  strongly suggesting that the state \fr{initial}, with $K(\theta)$ given by Eq. \fr{k_function}, is  at least a good approximation of the initial state after the quench described above for any value of $m_0$. 

In the following, we don't enter into the discussion of how the state \fr{initial} is prepared and address the quench problem where the initial condition is chosen to be given by Eq. \fr{initial}.

We stress that $K(\theta)$ defined above satisfies the \emph{boundary reflection equation}, namely
\be
K(\theta) = S(2 \theta) K(-\theta)\,.
\ee
As we will see later, in general it is important to require that, for large values of the rapidity, the function $K(\theta)$ is exponentially suppressed. This is true for the function $K(\theta)$ defined in Eq. \fr{k_function}, which satisfies
\be
K(\theta) \sim e^{-2\theta}\,,\qquad\text{for}\qquad\theta \gg 1\,.
\ee
This condition makes the state \fr{initial} well defined for a time-evolution problem.

\section{The post-quench stationary state}\label{sec:steady_state}

\subsection{The quench action method}\label{sec:QAM}

The quench action method (or representative eigenstate approach) is a recently developed analytical approach to study quantum quenches based on integrability. It was introduced in Ref.~\cite{ce-13} and so far has been mainly applied in the study of non-relativistic integrable models. Within the framework of quantum field theories, the QAM has been first employed in \cite{bertini}, where quantum quenches in the sine-Gordon model were investigated. In this section we will briefly review this approach, referring to the literature for a more detailed treatment \cite{ce-13}.

We begin our discussion by considering a theory defined on a finite system of length $L$ and denote with $|\Psi_0\rangle_L$ the initial state. The idea behind the QAM is to identify a representative eigenstate of the post-quench Hamiltonian that captures, in the thermodynamic limit, the stationary local properties of the systems at late times after the quench. More precisely, denoting by $\mathcal{O}(x)$ a generic local observable, the representative eigenstate $|\Phi\rangle_L$ is chosen in such a way that
\be
\label{Eq:REAtimedep}
\lim_{L\to\infty}\frac{\tensor*[_{L}]{\braket{\Psi_0|\mathcal{O}(x,t)|\Psi_0}}{_{L}}}{\tensor*[_{L}]{\braket{\Psi_0|\Psi_0}}{_{L}}}=\lim_{L\to\infty}\frac{1}{2}\left[\frac{\tensor*[_{L}]{\braket{\Psi_0|\mathcal{O}(x,t)|\Phi}}{_{L}}}{\tensor*[_{L}]{\braket{\Psi_0|\Phi}}{_{L}}}+\frac{\tensor*[_{L}]{\braket{\Phi|\mathcal{O}(x,t)|\Psi_0}}{_{L}}}{\tensor*[_{L}]{\braket{\Phi|\Psi_0}}{_{L}}}\right],
\ee
from which it is possible to derive
\be
\lim_{t\to\infty}\lim_{L\to\infty}\frac{\tensor*[_{L}]{\braket{\Psi_0|\mathcal{O}(x,t)|\Psi_0}}{_{L}}}{\tensor*[_{L}]{\braket{\Psi_0|\Psi_0}}{_{L}}}=\lim_{L\to\infty}\frac{\tensor*[_{L}]{\braket{\Phi|\mathcal{O}(x)|\Phi}}{_{L}}}{\tensor*[_{L}]{\braket{\Phi|\Phi}}{_{L}}}.
\label{Eq:infinitetimelimit}
\ee
Here we have used the standard notation $\mathcal{O}(x,t)$ for a time evolved operator within the Heisenberg picture, with $\mathcal{O}(x,0)=\mathcal{O}(x)$. Note that the existence of such a representative eigenstate is a highly non-trivial result \cite{ce-13}.

As we have seen in the previous section, each eigenstate of the system is characterized, in the thermodynamic limit, by the corresponding quasi-particle and hole rapidity distributions $\rho(\theta)$, $\rho^{h}(\theta)$. They satisfy the equations \fr{thermo_bethe_eq} but are otherwise undetermined. The representative eigenstate $|\Phi\rangle_L$ is then specified by an additional functional relation between $\rho(\theta)$ and $\rho^{h}(\theta)$, which is obtained as the saddle-point condition of an appropriate functional $S_{QA}[\rho]$ \cite{ce-13}, namely
\be
\frac{\delta}{\delta \rho}S_{QA}[\rho]=0.
\label{saddle_point}
\ee
The functional $S_{QA}$ (the so called quench action) is defined as
\be
S_{QA}[\rho]=2S[\rho]-S_{YY}[\rho].
\label{qa}
\ee
Here
\be
S[\rho]=-\lim_{L\to\infty}{\rm Re}\ln \tensor*[_{L}]{\braket{\Psi_0|\rho}}{_L}\,,
\label{overlap}
\ee
where $\ket{\rho}_L$ is an eigenstate of the finite volume Hamiltonian corresponding, in the thermodynamic limit, to the root density $\rho(\theta)$ (the hole rapidity distribution $\rho^h(\theta)$ is then uniquely fixed by Eq.~\eqref{thermo_bethe_eq}). The functional $S_{YY}[\rho]$ is the Yang-Yang entropy of the distribution $\rho(\theta)$. As we will discuss in the following section, the initial state \fr{initial} belongs to the sector of the post-quench Hilbert space corresponding to parity-symmetric (or parity-invariant) Bethe states. In this case, the Yang-Yang entropy appearing in \fr{qa} reads
\be
S_{YY}[\rho]=\frac{L}{2}\int_{-\infty}^{\infty}d\theta\left[\left(\rho+\rho^h\right)\ln\left(\rho+\rho^h\right)-\rho\ln\rho-\rho^h\ln\rho^h\right],
\ee
where the overall factor $1/2$ stems from the restriction to the parity-symmetric sector of the post-quench Hilbert space \cite{dwbc-14,wdbf-14, pmwk-14, pce-15}. Given the saddle-point condition \fr{saddle_point}, the representative eigenstate is usually denoted with $|\rho_{\rm sp}\rangle_L$. Note that while the Yang-Yang entropy has an integral representation which is independent of the initial state, the functional $S[\rho]$, and hence $S_{QA}[\rho]$ can be explicitly written down only when an expression for the overlaps between the initial state and the eigenstates of the post-quench Hamiltonian is known. In the following section we show that these overlaps can be easily obtained for initial states of the form \fr{initial}. 

\subsection{The saddle-point equations and the steady state}\label{sec:QAM_equations}

In this section we derive the saddle-point equations for initial states of the form \fr{initial} and we explicitly solve them for the mass and interaction quenches discussed in the previous section, cf. Eq. \fr{k_function}.

In Ref.~\cite{bertini} the quench action approach was first applied within the framework of quantum field theories. There, in order to obtain the overlaps between the initial state and the excited states of the system, the authors considered a finite-volume regularization of the theory, as it was systematically investigated in Refs.~\cite{pt-08, kp-10}. In the following we will proceed along the lines of Ref.~\cite{bertini}.

We start by considering the regularization of the initial states of the form \fr{initial} in a large finite volume, as obtained in Ref.~\cite{kp-10}. This reads
\be
|\Psi_0\rangle_{L}=\frac{1}{\mathcal{Z}(L)}\sum_{N=0}^{+\infty}\sum_{0<I_1<\ldots<I_N}\mathcal{N}_{2N}(\theta_1,\ldots,\theta_N)\mathcal{K}(\theta_1,\ldots,\theta_N)|-I_1,I_1,\ldots,-I_N,I_N\rangle,
\label{finite_sys}
\ee
where
\bea
\mathcal{K}(\theta_1,\ldots,\theta_N)&\equiv& \prod_{j=1}^{N}K(\theta_j),\\
\mathcal{N}_{2N}(\theta_1,\ldots,\theta_N)&\equiv&\frac{\sqrt{\rho_{2N}\left(-\theta_1,\theta_1,\ldots, -\theta_N,\theta_N\right)}}{\overline{\rho}_N(\theta_1,\ldots,\theta_N)},
\eea
while $\mathcal{Z}(L)$ is the global normalization constant (which depends on the system size $L$). Here, we have used the definitions
\bea
\rho_N(\theta_1,\ldots, \theta_N) &=& \det \mathcal{J}_N(\theta_1,\ldots,\theta_N),\qquad \mathcal{J}_N(\theta_1,\ldots,\theta_N)_{ij}=\partial_{\theta_j}Q_i\left(\theta_1,\ldots,\theta_N\right),\\
\overline{\rho}_N(\theta_1,\ldots, \theta_N) &=& \det \overline{\mathcal{J}}_N(\theta_1,\ldots,\theta_N),\qquad \overline{\mathcal{J}}_N(\theta_1,\ldots,\theta_N)_{ij}=\partial_{\theta_j}\overline{Q}_i\left(\theta_1,\ldots,\theta_N\right),\
\eea
where $\mathcal{Q}_j$, $\overline{\mathcal{Q}}_j$ are defined in Eqs. \fr{q_function}, \fr{q_bar_function}. The sum in \fr{finite_sys} is over the set of positive quantum numbers $\{I_j\}_{j=1}^{N}$, which label the eigenstates of the parity symmetric sector of the Hilbert space
\be
|-I_1,I_1,\ldots,-I_N,I_N\rangle.
\label{l_eigenstates}
\ee
The eigenstates \fr{l_eigenstates} are chosen to be of unitary norm. The sets of positive rapidities $\{\theta_j\}_{j=1}^{N}$ appearing in the different terms of the sum \fr{finite_sys} are defined as follows: for any given set of positive quantum numbers $\{I_j\}_{j=1}^{N}$, they are the positive solutions of the Bethe equations \fr{bethe_eq} using the $2N$ quantum numbers $\{-I_j\}_{j=1}^{N}\cup \{I_j\}_{j=1}^{N}$.
 
Note that for finite $L$ the state \fr{finite_sys} has finite norm (namely, $\mathcal{Z}(L)<\infty$) provided that the function $K(\theta)$ goes to zero sufficiently fast. It is useful to remind here that the expression \fr{finite_sys} was derived in Ref.~\cite{kp-10} by imposing that the expectation value of local operators in the state $|\psi_0\rangle_L$ must reproduce, when $L\to\infty$, the known infinite volume values up to exponentially vanishing corrections in the system size. 

\begin{figure}
\includegraphics[scale=0.85]{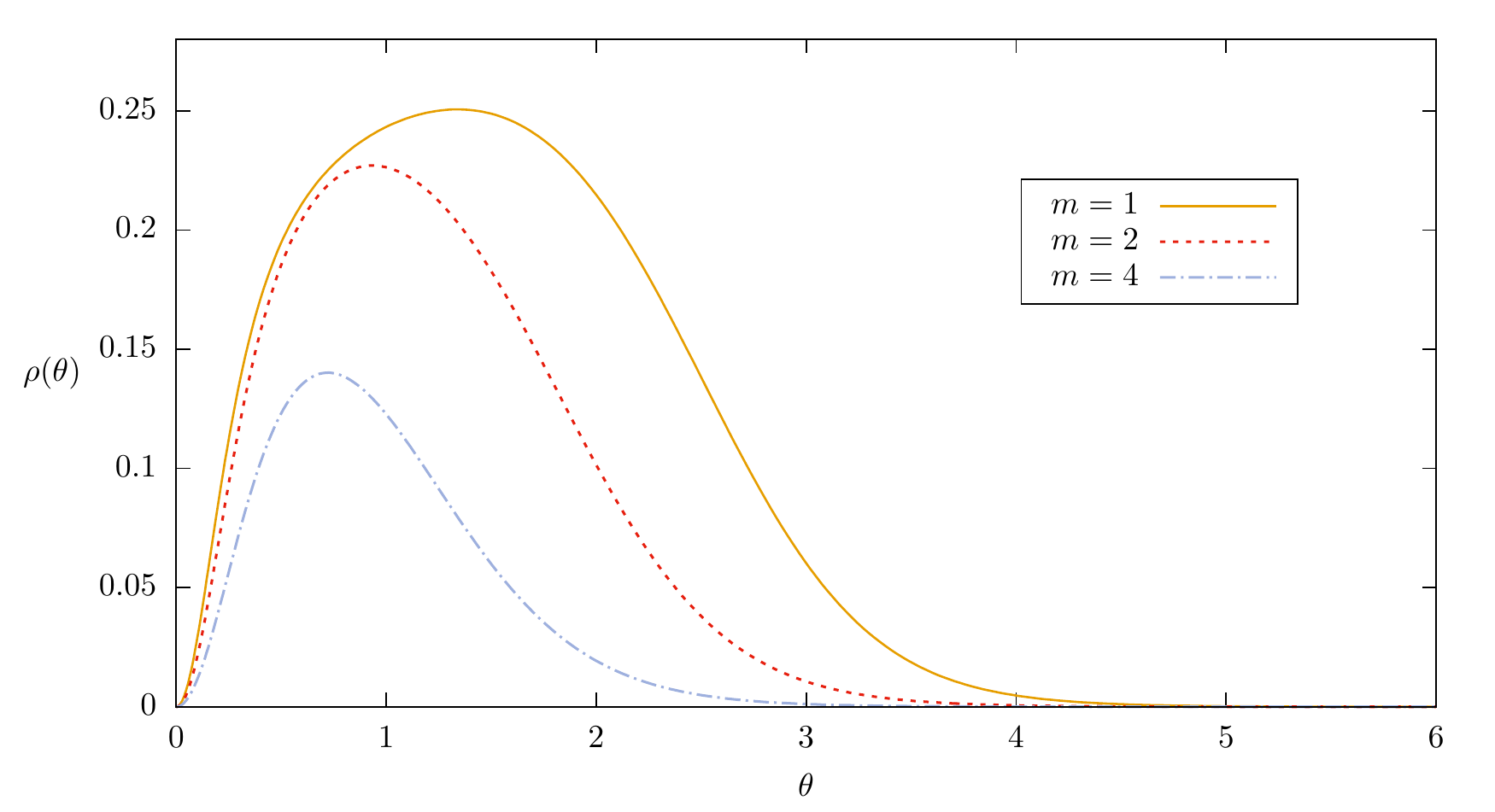}
\caption{(Color online) Rapidity distributions $\rho(\theta)$ of the post-quench stationary state for different masses $m$ of the post-quench Hamiltonian. The plot corresponds to the case of initial mass $m_0=10$ and final coupling $\alpha=0.5$ (here we set $c=1$). The different curves correspond respectively to final mass $m=4$ (blue line), $m=2$ (red line) and $m=1$ (orange line). Note that for decreasing final mass we have a larger density of excitations, namely increasing $\Delta m=m_0-m$ a larger number of particles is produced after the quench.}
\label{fig_saddle}
\end{figure}

From the representation \fr{finite_sys} it is now easy to derive the leading term of the logarithm of the overlap \fr{overlap}. First, note that for $L\to\infty$, we have
\be
\mathcal{N}_{2N}(\theta_1,\ldots,\theta_N)\sim 1+\mathcal{O}\left(\frac{1}{L}\right),
\label{determinant_ratio}
\ee
namely $\mathcal{N}_{2N}$ gives only sub-leading contributions to the overlap and can be neglected in the following. Eq. \fr{determinant_ratio} follows directly by the results of \cite{pozsgay-10_II} and it is analogous to the situation encountered in the Lieb-Liniger model in \cite{dwbc-14}. Then, the final result simply reads
\be
S[\rho]=-L\int_{0}^{\infty}{d}\theta \rho(\theta)\log |K(\theta)|+C,
\ee
where $C$ is a constant generated by the global (unknown) normalization $\mathcal{Z}(L)$ in \fr{finite_sys}.

We can then finally explicitly write down the saddle-point equations discussed in the previous section, cf. Eq. \fr{saddle_point}. After straightforward calculations we finally obtain
\be
\ln \eta_{\rm sp}(\theta)=-\ln|K(\theta)|^2-\int_{-\infty}^{\infty}\frac{{\rm d}\theta'}{2\pi}\varphi_{\alpha}(\theta-\theta')\ln\left(1+\frac{1}{\eta_{\rm sp}(\theta')}\right).
\label{e_saddle_point}
\ee
where $\varphi_{\alpha}(\theta)$ is defined in \fr{Eq:varphi}.

Note that we have no chemical potential in this equation, as opposed to the non relativistic cases~\cite{wdbf-14, pmwk-14,ac-15, dwbc-14, pce-15, bucciantini-15}. 
There, one considers as initial state the ground state of the pre-quench Hamiltonian with a fixed density $D$ of particles. As a consequence, a term of the form
\be
h\left(\int\rho(\theta){\rm d}\theta-D\right)
\ee
has to be added to the action \eqref{qa}, where $h$ is a Lagrange multiplier.

Eqs. \eqref{thermo_bethe_eq} and \eqref{e_saddle_point} completely determine the saddle-point distribution and hence the post-quench stationary state. They can be numerically solved using standard techniques. In Fig. \ref{fig_saddle} the rapidity distribution for the saddle-point is shown for different values of the quench parameters.

In the next section we will focus on determining the stationary values of one-point correlation functions, explicitly evaluating the r.h.s. of Eq.~\fr{Eq:infinitetimelimit}. In particular, we present a formula for the exact computation of one-point functions of generic vertex operators $:e^{\alpha \phi(x)}:$ in any excited state described by a pair of root densities $\rho(\theta),\, \rho^h(\theta)$ satisfying the Bethe equations \fr{thermo_bethe_eq}. 

\section{One point correlation functions}\label{sec:one_point}

In this section we address the computation of one-point correlation functions in the post-quench stationary state. Specifically we consider the expectation value of the \emph{vertex operators}, defined as 
\be
\label{Eq:vertex}
\Phi_{k}(x)\equiv\, :e^{k g \phi(x)}:\,\qquad k\in\mathbb{R}\,,
\ee
where $g$ is the same parameter appearing in the Hamiltonian \fr{hamiltonian}. These operators are particularly interesting because they provide the generating function for the powers of the field $:\phi^n(x):$, with $n$ positive integer. In the following, after reviewing the LeClair-Mussardo formalism, we proceed by presenting and discussing our formula for the expectation value of the vertex operators \fr{Eq:vertex}. 

Throughout this section we make use of the following convenient shorthand notation
\be
\braket{\mathcal{O}(x)}_\rho \equiv \lim_{L\rightarrow \infty} \frac{\tensor*[_{L}]{\braket{\rho|\mathcal{O}(x)|\rho}}{_{L}}}{\tensor*[_{L}]{\braket{\rho|\rho}}{_{L}}}\,,
\ee
where, as in the previous section, $\ket{\rho}_L$ is an eigenstate of the finite volume Hamiltonian corresponding, in the thermodynamic limit, to the root density $\rho(\theta)$.  

\subsection{The LeClair-Mussardo series}\label{sec:LM}

The LeClair-Mussardo series is a general method to compute one point functions of local operators on arbitrary excited states in integrable quantum field theories. It was first introduced for computing one point functions at finite temperature in Ref.~\cite{lm-99}. There, the authors argued that the expectation of a local operator ${\mathcal O}(x)$ on a thermal state at temperature $T$ is given by the following expansion 
\be
\label{Eq:LMF}
\braket{{\mathcal O}(x)}_{\rho_T}=\braket{0|O(0)|0}+\sum_{n=1}^{\infty}\frac{1}{n!}\int_{-\infty}^{\infty}\cdots\int_{-\infty}^{\infty}\prod_{i=1}^{n}\left(\frac{{\rm d}\theta_i}{2 \pi}\frac{1}{1+e^{\varepsilon_T(\theta_i)}}\right)\braket{\theta_n,\cdots,\theta_1|\mathcal{O}(0)|\theta_1,\cdots,\theta_n}_c\,,
\ee
where $\varepsilon_T(\theta)$ is the solution of the integral equation
\be
\label{Eq:TBA}
\varepsilon_T(\theta) =\frac{mc^2}{T}\cosh\theta-\int_{-\infty}^{\infty}\frac{{\rm d}\theta'}{2 \pi}\varphi_\alpha(\theta-\theta')\log[1+e^{-\varepsilon_T(\theta')}] \,,
\ee
and $\varphi_{\alpha}(\theta)$ is defined in Eq. (\ref{Eq:varphi}). By $\braket{\theta_n,\cdots,\theta_1|\mathcal{O}(0)|\theta_1,\cdots,\theta_n}_c$ we denote the \emph{connected part} of the diagonal $n$-particle \emph{form factor} of the operator ${\mathcal O}(x)$, which is defined as
\be 
\label{Eq:limit}
\braket{\theta_n,\cdots,\theta_1|\mathcal{O}(0)|\theta_1,\cdots,\theta_n}_c = \lim_{\{\zeta_i\}\rightarrow0^+}\braket{0|{\mathcal O}|\theta_1,\cdots,\theta_n,\theta_n- i \pi + i \zeta_n,\cdots,\theta_1- i \pi + i \zeta_1}\big|_{\textrm{finite part}}\,.
\ee
The prescription of taking the \emph{finite part} in the limit consists of neglecting all the divergent contributions in the form $1/\zeta_i^{p}$ with $p>0$ or $\zeta_i/\zeta_j$. 

In Ref.~\cite{pozsgay-11} it was shown that the LeClair-Mussardo series, initially conceived for expectation values at finite temperature, can be used for computing one point functions of arbitrary excited states. Specifically, given a macroscopic state described by the distributions $\rho$, $\rho^{h}$, one can directly apply Eq. \fr{Eq:LMF} with the simple substitution
\be
\varepsilon_T(\theta)\to\varepsilon(\theta)\equiv \log\left(\eta(\theta)\right)=\log \left(\frac{\rho^h(\theta)}{\rho(\theta)}\right).
\label{Eq:aux}
\ee
In particular, plugging into Eq.~\fr{Eq:LMF} the solution $\eta_\textrm{sp}(\theta)$ of the saddle point equation \fr{e_saddle_point} we immediately arrive at the following expression for the one-point functions on the post-quench stationary state 
\be
\label{Eq:LMFQA}
\braket{{\mathcal O}(x)}_{\rho_{\rm sp}}=\braket{0|O(0)|0}+\sum_{n=1}^{\infty}\frac{1}{n!}\int_{-\infty}^{\infty}\cdots\int_{-\infty}^{\infty}\prod_{i=1}^{n}\left(\frac{{\rm d}\theta_i}{2 \pi}\frac{1}{1+\eta_{\rm sp}(\theta_i)}\right)\braket{\theta_n,\cdots,\theta_1|\mathcal{O}(0)|\theta_1,\cdots,\theta_n}_c\,.
\ee
This result for the long-times post-quench stationary value of one-point correlation functions was first obtained in Refs.~\cite{fioretto-10, pozsgay-11}, where the same equation for $\eta_{\rm sp}(\theta)$ was derived by different methods. In particular, as it was argued in Refs.~[\onlinecite{fioretto-10,mussardo}], Eqs. \fr{e_saddle_point} and \fr{Eq:LMFQA} match the predictions obtained by the GGE constructed with the conserved rapidity occupation numbers.

Despite its conceptual importance, from a practical point of view Eq. \fr{Eq:LMFQA} can be numerically evaluated only in the ``small quench'' limit, i.e. when the function 
\be
r_{\textrm{sp}}(\theta)=\frac{1}{1+\eta_{\rm sp}(\theta)},
\ee
is uniformly small. Note that in the specific mass quench described in section \ref{sec:initial} this corresponds to choosing $m_0\sim m$, cf. Eq. \fr{k_function}. In this case, terms corresponding to increasing values of $n$ in the r.h.s. of Eq. \fr{Eq:LMFQA} are rapidly suppressed by higher powers of $r_{\rm sp}(\theta)$ and the series can be truncated to the first few terms. Away from this regime one needs in general to compute a large number of terms, each involving a non-trivial multidimensional integral. Moreover, the actual expression of the connected form factors gets progressively more unwieldy as $n$ increases. These complications pose serious limitations to the numerical evaluation of the LeClair-Mussardo series.

In the next section we propose a formula which allows to efficiently compute the exact expectation value of the vertex operators $\Phi_{k}(x)$ on arbitrary excited states, overcoming the difficulties discussed before. As we will explain, our formula is based on recent results by Negro and Smirnov for thermal expectation values in the sinh-Gordon model \cite{ns-13,negro-14}.

\subsection{Expectation value of vertex operators: exact formula}\label{sec:NS}

We now present the main result of this section. Given a state described by the macroscopic rapidity distributions $\rho(\theta)$, $\rho^h(\theta)$ [related by Eq.~(\ref{thermo_bethe_eq})] we propose the following expression for the ratio of expectation values of the vertex operators $\Phi_k(x)$ defined in Eq. \fr{Eq:vertex}
\be
\frac{\left\langle \Phi_{k+1}(x)\right\rangle_{\rho}}{\langle \Phi_{k}(x) \rangle_{\rho}}=1+\frac{2\sin\left(\pi \alpha (2 k +1)\right)}{\pi}\int_{-\infty}^{\infty} \!{\rm d}\theta\frac{e^\theta}{1+\eta(\theta)} p_k(\theta)\,,
\label{Eq:main}
\ee
where $\eta(\theta)$ is defined in Eq. \fr{def_eta}. Here we introduced the function $p_k(\theta)$ defined by the following integral equation
\begin{align}
&p_k(\theta)=e^{-\theta}+\int_{-\infty}^{\infty} \!{\rm d}\mu\frac{1}{1+\eta(\mu)} \chi_k(\theta-\mu)p_k(\mu)\,,
\label{Eq:p}
\end{align}
where
\be
\label{def_chi}
\chi_{k}(\theta)\equiv\frac{i}{2 \pi}\left(\frac{e^{-i 2 k \alpha \pi}}{\sinh\left(\theta+i\pi \alpha\right)}-\frac{e^{i2 k \alpha \pi}}{\sinh\left(\theta-i{\pi}\alpha \right)}\right)\,.
\ee
Note that Eq. \fr{Eq:main} provides an expression for the {\it ratios} of expectation values of vertex operators. However, if the coupling $\alpha$ [defined in Eq. \fr{Eq:alpha}] is irrational, Eq. \fr{Eq:main} can easily be used to obtain the expectation values $\braket{\Phi_k(x)}_{\rho}$ for arbitrary $k$, as we now explain. Note that it is not a limitation to require that $\alpha$ is an irrational number. Indeed, every real number can be approximated arbitrarily well with an irrational one.

The procedure to obtain the values $\langle \Phi_k(x)\rangle_{\rho}$ out of the ratios \fr{Eq:main} can be explained as follows
\begin{itemize}
\item[(i)] First, note $\braket{\Phi_0(x)}_{\rho}=1$, as $\Phi_0(x)$ is just the identity operator. Using this, and repeatedly applying \fr{Eq:main}, we can construct the sequence 
\be
\{\braket{\Phi_n(x)}_{\rho}\}_{n\in \mathbb{N}}\,.
\label{Eq:sequence}
\ee
\item[(ii)] Second, observe that the expectation value $\braket{\Phi_k(x)}_{\rho}$ is $\frac{1}{\alpha}$-periodic in $k$, \emph{i.e.} $\braket{\Phi_k(x)}_{\rho}=\braket{\Phi_{k+{\alpha}^{-1}}(x)}_{\rho}$. This can be seen employing the LeClair-Mussardo series representation of $\braket{\Phi_k(x)}_{\rho}$ and using that the exact expression of the form-factors $F_{2n}^k(\theta_1,\ldots,\theta_{2n})=\braket{0|\Phi_k(0)|\theta_1,\ldots,\theta_{2n}}$, found in \cite{KM:FF}, shows this periodicity in $k$. We can then use this periodicity to fold back the sequence \fr{Eq:sequence} on top of the interval $[0,{\alpha}^{-1}]$ by assigning to every $n>{\alpha}^{-1}$ the real number 
\be
\label{Eq:folding}
\tilde n = n - \frac{m}{\alpha}\,,
\ee
with $m\in \mathbb{N}$ chosen in such a way that $\tilde n \in [0,{\alpha}^{-1}]$. Because of the periodicity we have $\braket{\Phi_n(x)}_{\rho}=\braket{\Phi_{\tilde n}(x)}_{\rho}$.
\item[(iii)] Finally, it is easy to show that, for irrational $\alpha$, every term of the sequence \fr{Eq:sequence} corresponds, under the folding procedure, to a different value in the interval $[0,{\alpha}^{-1}]$. This simply follows from the fact that $\frac{n}{\alpha}\notin \mathbb{N}$ for $n\in \mathbb{N}$. As a consequence, our sequence densely covers the interval $[0,{\alpha}^{-1}]$ under the folding procedure \fr{Eq:folding}.
\end{itemize}

\begin{figure}
\includegraphics[scale=0.9]{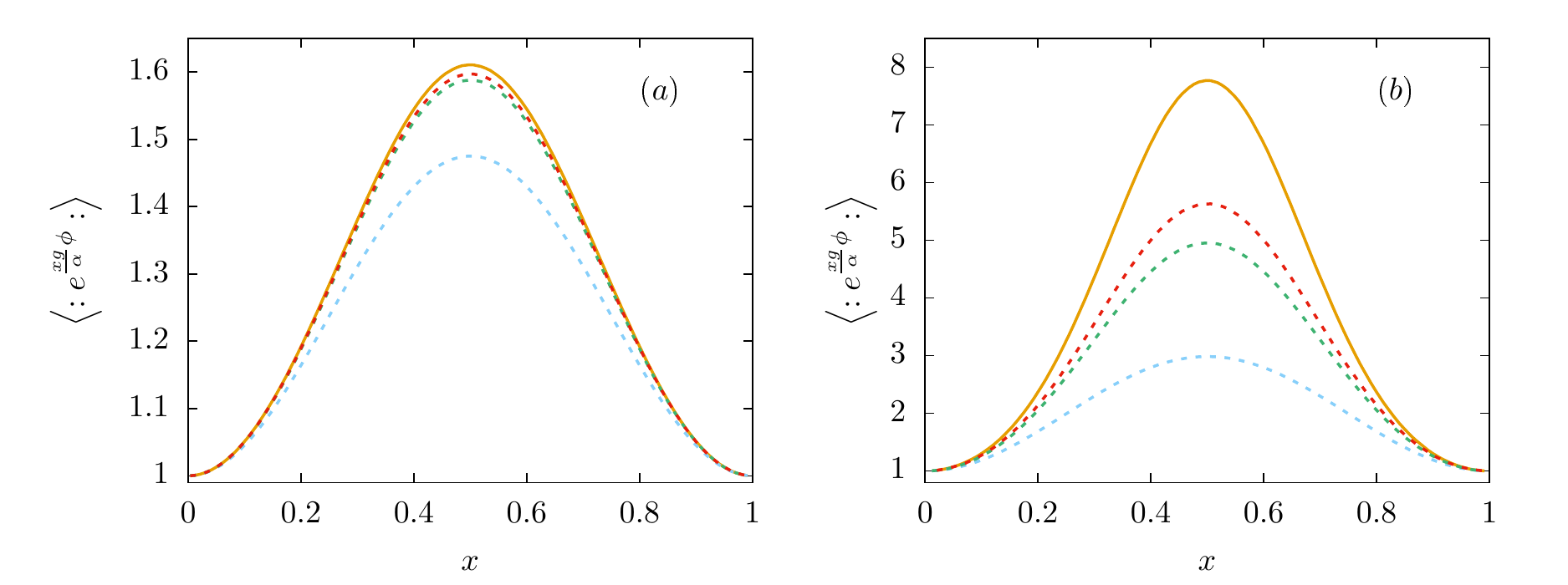}
\caption{(Color on-line) Expectation value of the vertex operator $\exp[xg\phi(0)/\alpha]$ as a function of the parameter $x$. The orange solid lines correspond to the exact values computed using Eq. \fr{Eq:main}, while the dashed lines correspond to the approximated values obtained by the LeClair-Mussardo series \fr{Eq:LMFQA} truncated at the order $n=1$ (blue dashed lines), $n=2$ (green dashed lines) and $n=3$ (red dashed lines). The plots correspond to $\alpha=1/\sqrt{2}$ while the ratio of the initial mass $m_0$ and the final mass $m$ is: $(a)$ $m/m_0=0.3$, $(b)$ $m/m_0=0.08$. } 
\label{Fig:vertex}
\end{figure}

Importantly, we note that formulae \fr{Eq:main} and \fr{Eq:p} are extremely convenient for the numerical evaluation at arbitrary values of the ratio $\eta(\theta)$ defined in Eq. \fr{def_eta}. Combined with our previous analysis by the QAM, in particular, they allow one to compute the post-quench stationary expectation value of vertex operators after initializing the system in any initial state of the form \fr{initial}. In Fig. \ref{Fig:vertex} we show the application of Eq.~\fr{Eq:main} to the quench problem discussed in section \ref{sec:initial}, cf. Eq.~\fr{k_function}. We see that one can easily obtain explicit numerical results in non-perturbative regimes also, where the first few terms in the LeClair-Mussardo series do not provide a good approximation.

The remaining part of this section is devoted to explain the origin of the formula \fr{Eq:main} and to check its predictions in cases where explicit independent results can be found.

\subsection{Derivation of Eq. \fr{Eq:main}: the Negro-Smirnov formula}

The starting point for the derivation of Eq. \fr{Eq:main} is a recent formula presented by Negro and Smirnov for the ratio of thermal expectation values of vertex operators  \cite{ns-13, negro-14}. Within our notations, it reads 
\be
\frac{\left\langle \Phi_{k+1}(x)\right\rangle_{\rho_T}}{\langle \Phi_{k}(x) \rangle_{\rho_T}}=1+\frac{2\sin\left(\pi \alpha (2 k +1)\right)}{\pi}\left(\,\, \int_{-\infty}^{\infty}{\rm d}\theta\, \frac{1}{1+e^{\varepsilon_T(\theta)}}+\int_{-\infty}^{\infty}{\rm d}\mu \,\frac{e^{\mu}}{1+e^{\varepsilon_T(\mu)}} \int_{-\infty}^{\infty}{\rm d}\lambda\, R_T(\mu,\lambda|k)\,\frac{e^{-\lambda}}{1+e^{\varepsilon_T(\lambda)}}\right)\,,\label{Eq:start}
\ee
where, as before, the operators $\Phi_k(x)$ are defined in Eq. \fr{Eq:vertex}, and $\varepsilon_T(\theta)$ is the solution of the TBA equation \fr{Eq:TBA}. The function $R_T(\theta,\theta'|k)$ is a two-variable function defined as the solution of the following integral equation  
\be
R_T(\theta,\theta'|k)-\int_{-\infty}^{\infty}{\rm\ d}\mu\,\frac{1}{1+e^{\varepsilon_T(\mu)}}\,\chi_{k}(\theta-\mu)R_T(\mu,\theta'|k)=\chi_{k}(\theta-\theta')\,,
\label{Eq:Rdressed_thermal}
\ee
where $\chi_k(\theta)$ is defined in \fr{def_chi}. Using fermionic basis techniques \cite{fermionicbasis}, formula \fr{Eq:start} was first conjectured in Ref.~\cite{ns-13}, where its validity is checked against previously known limiting-case expressions. Further convincing numerical evidence in favour of the validity of Eq. \fr{Eq:start} was subsequently given in Ref.\,\cite{negro-14}.

At low temperatures, a perturbative expansion of the r.h.s. of Eq. \fr{Eq:start} is possible since the function
\be
r_T(\theta)\equiv \frac{1}{1+e^{\varepsilon_T(\theta)}}\,
\ee
is easily seen to be exponentially vanishing for $T\to 0$. In Ref.~\cite{ns-13} it was shown that up to the third order in $r_T(\theta)$, the resulting expression matches the analogous expansion of the l.h.s. obtained by employing the LeClair-Mussardo series \fr{Eq:LMF}. The fact that $\varepsilon_T(\theta)$ is the solution of the TBA equation \fr{Eq:TBA} is never used in this expansion. Indeed, we explicitly verified that, up to order $r(\theta)^3$, the expansions of the r.h.s. of Eq. \fr{Eq:start} and its l.h.s. (obtained starting from the LeClair-Mussardo series) are identical keeping $r(\theta)$ as a functional variable.

In Appendix~\ref{App:CFF} we report the first three form factors of the vertex operator $\Phi_k(x)$. One can see that their expression gets progressively intricate as their order increases. This reflects in the difficulty to verify that the expansions of the l.h.s. and r.h.s. of Eq. \fr{Eq:start} are equal for higher orders. Nevertheless, it natural to assume that this is the case, which immediately leads to
\be
\frac{\left\langle \Phi_{k+1}(x)\right\rangle_{\rho}}{\langle \Phi_{k}(x) \rangle_{\rho}}=1+\frac{2\sin\left(\pi \alpha (2 k +1)\right)}{\pi}\left(\,\, \int_{-\infty}^{\infty}{\rm d}\theta\, \frac{1}{1+\eta(\theta)}+\int_{-\infty}^{\infty}{\rm d}\mu \,\frac{e^\mu}{1+\eta(\mu)} \int_{-\infty}^{\infty}{\rm d}\lambda\, \,\frac{e^{-\lambda}}{1+\eta(\lambda)}R(\mu,\lambda|k)\right)\,,\label{Eq:step1}
\ee
where as usual $\eta(\theta)$ is given in Eq. \fr{def_eta} and where $R(\theta,\theta'|k)$ solves 
\be
R(\theta,\theta'|k)-\int_{-\infty}^{\infty}{\rm\ d}\mu\, \frac{1}{1+\eta(\mu)}\,\chi_{k}(\theta-\mu)R(\mu,\theta'|k)=\chi_{k}(\theta-\theta')\,.
\label{Eq:Rdressed}
\ee
Here, it is worth stressing that Eq. \fr{Eq:step1} is simply obtained by Eq. \fr{Eq:start} by the substitution \fr{Eq:aux}. This is motivated, although not rigorously proven, by the validity of such a substitution when applied to the LeClair-Mussardo series as established in Ref.~\cite{pozsgay-11}. 

Formula \fr{Eq:step1} can now be simplified. We introduce the function $p_k(\theta)$ defined by 
\be
p_k(\theta):=e^{-\theta}+\int_{-\infty}^{\infty}\!{\rm d}\lambda\frac{e^{-\lambda}}{1+\eta(\lambda)}R(\theta,\lambda|k),
\label{Eq:new_p}
\ee
so that
\be
\int_{-\infty}^{\infty}{\rm d}\theta\, \frac{1}{1+\eta(\theta)}+\int_{-\infty}^{\infty}{\rm d}\mu \,\frac{e^\mu}{1+\eta(\mu)} \int_{-\infty}^{\infty}{\rm d}\lambda\, \,\frac{e^{-\lambda}}{1+\eta(\lambda)}R(\mu,\lambda|k)=\int_{-\infty}^{\infty} \!{\rm d}\theta	\frac{ e^\theta}{1+\eta(\theta)} p_k(\theta)\,.
\label{Eq:step2}
\ee
In order to derive Eq. \fr{Eq:main}, now we have to show that the function $p_k(\theta)$ satisfies the linear integral equation \fr{Eq:p}. This is simply achieved by multiplying both sides of Eq. \fr{Eq:Rdressed} by 
\be
\frac{e^{-\theta'}}{1+\eta(\theta')},
\ee
and integrating in the variable $\theta'$. Using the definition \fr{Eq:new_p} we straightforwardly arrive at Eq. \fr{Eq:p}.

\subsection{Comparison with known results}\label{sec:Comp}

In this section we show that formula \fr{Eq:main} is consistent with independent results obtained by the LeClair-Mussardo series and the Feynman-Hellmann theorem.

\subsubsection{Comparison with the LeClair-Mussardo series}

As we already mentioned, in order to prove the validity of our formula, one should show that the r.h.s. of Eq. \fr{Eq:main} is the re-summation of the series expansion of the ratio
\be
\frac{\braket{\Phi_{k+1}(0)}_{\rho}}{\braket{\Phi_{k}(0)}_{\rho}}\,,
\label{Eq:exprat}
\ee
in powers of the functional parameter
\be
r(\theta)=\frac{1}{1+\eta(\theta)},
\label{eq:r_parameter}
\ee
where $\eta(\theta)$ is defined in Eq. \fr{def_eta}. This expansion is obtained by writing both the numerator and denominator of the fraction in Eq. \fr{Eq:exprat} as a LeClair-Mussardo series and then compute the resulting series expansion for their ratio. The main difficulty of this calculation lies on the fact that the expression of the connected $n$-particle form factor is increasingly involved with $n$ as reported in appendix \ref{App:CFF}. Accordingly, one can analytically compute only the first few terms. In particular, we explicitly obtained an expansion in $r(\theta)$ of the ratio \fr{Eq:exprat} to the third order. An analogous expansion in $r(\theta)$ can be obtain for the r.h.s. of Eq. \fr{Eq:main} by a perturbative iterative solution of Eq. \fr{Eq:p}. We analytically verified that the two expansions are identical up to the third order.

\begin{figure}
\includegraphics[scale=0.85]{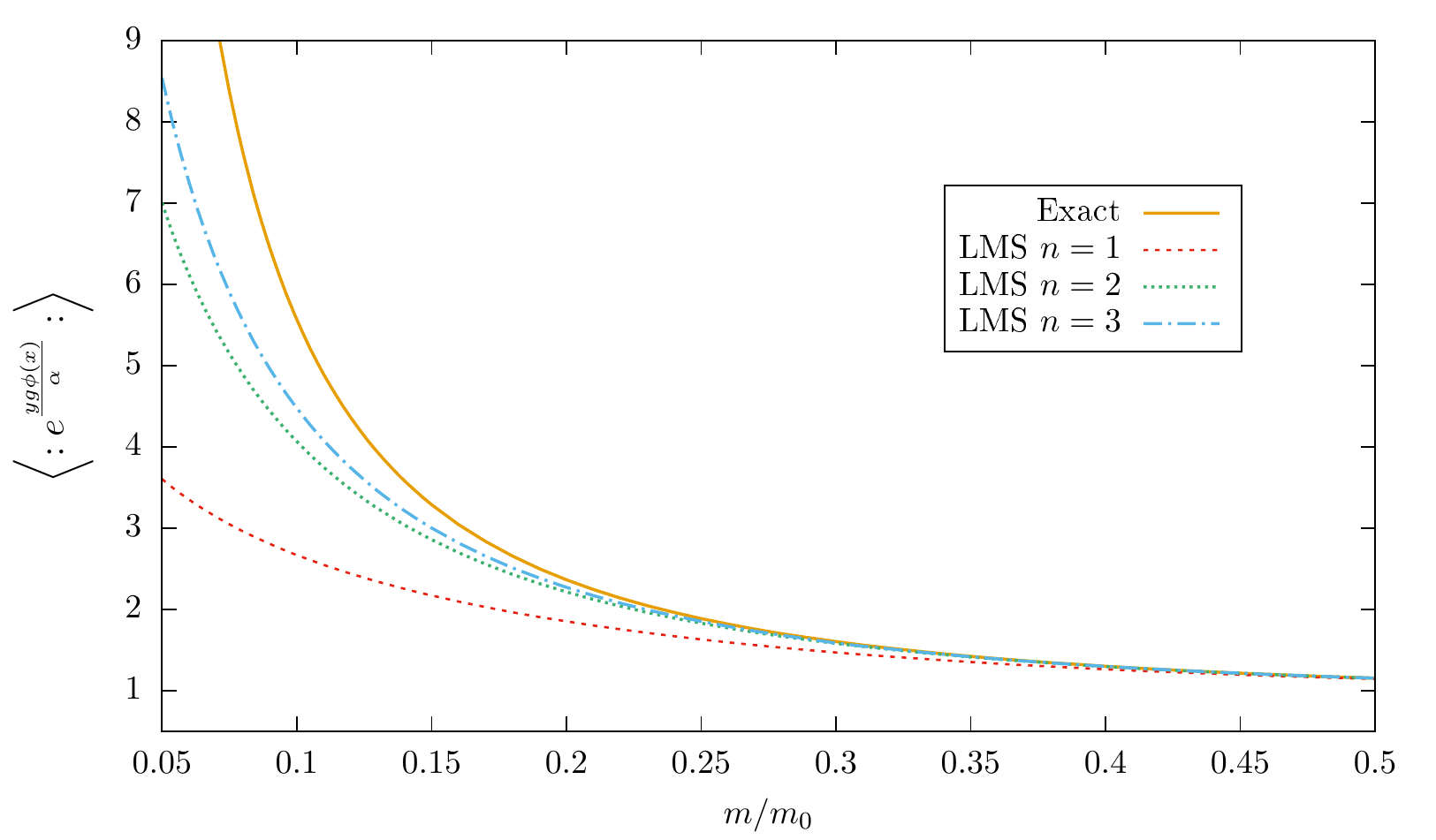}
\caption{(Color on-line) Post-quench stationary expectation value of a generic vertex operator $\exp[y g\phi(0)/\alpha]$ as a function of the ratio $m/m_0$ of the final and initial masses, for the quench described in section \ref{sec:initial}. Here we chose $y=0.53$, while the post-quench coupling is $\alpha=1/\sqrt{2}$. The predictions of the LeClair-Mussardo series truncated at order $n=1,2,3$ (dashed lines) are compared with the numerical evaluation of formula \fr{Eq:main} (solid orange line).}
\label{Fig:compmass}
\end{figure}

In Fig.~\ref{Fig:compmass} we report the comparison between the numerical computation of formula \fr{Eq:main} and the LeClair-Mussardo series truncated at order $n=1,2,3$ for the mass and interaction quench discussed in section \ref{sec:initial}, cf. Eq. \fr{initial}, \fr{k_function}. Specifically, the expectation value of the vertex operator $\Phi_k(x)$ on the post-quench stationary state is displayed as a function of the final mass $m$. We see that for small quenches, namely $m/m_0\sim 1$ the first terms of the LeClair-Mussardo series provide an excellent approximation to the exact value, while increasing the difference between the initial and the final masses, more and more terms should be kept.

\subsubsection{The Feynman--Hellmann theorem}\label{sec:FH}

The Feynman-Hellmann theorem provides an independent way to the compute the expectation value on arbitrary excited states of the particular vertex operator $\Phi_{1}(0)$, namely in the special case of Eq. \fr{Eq:vertex} corresponding to $k=1$. In this section we verify that formula \fr{Eq:main} recovers the result of the Feynman-Helmann theorem in this special case, hence providing further non-trivial evidence for the validity of our formula.

We consider our system confined in a large finite volume $L$, and assume it to be in the state 
\be
\ket{\rho}_{N,L}=\ket{\theta_1,\cdots,\theta_N}_L\,.
\label{Eq:state}
\ee 
The state $\ket{\rho}_{N,L}$ is an eigenstate of the finite volume Hamiltonian $H_L$, \emph{i.e.} its rapidities satisfy \fr{bethe_eq}. For later reference we observe that the state \fr{Eq:state} can always be chosen to be an eigenstate of the 
$\mathbb{Z}_2$ symmetry \fr{Eq:Z2}.
 
As usual, we assume that the state $\ket{\rho}_{N,L}$ is described by the distribution $\rho(\theta)$ when the thermodynamic limit is considered. The energy of the state $\ket{\rho}_{N,L}$ is given by   
\be
E^{\rho}_{N,L}= L \mathcal{E}_0 + \sum_{i=1}^{N} m \cosh \theta_i\,.
\ee
Here sub-leading corrections in the system size have been neglected and $\mathcal{E}_0$ is the bulk energy density of the infinite volume ground state, given by \cite{DdV:freeenergy} 
\be
\mathcal{E}_0= \frac{m^2}{8 \sin \pi \alpha}\,,
\ee
where here and in the rest of this section we have set for convenience the velocity of light equal to unity, namely $c=1$. 

In this setting, the expectation value $\tensor*[_{N,L}]{\braket{\rho|:\!\cosh g \phi(x)\!:|\rho}}{_{N,L}}$ can be computed by means of Feynman-Hellmann theorem  
\be
\lim_{L\rightarrow\infty}\frac{1}{L}\tensor*[_{N,L}]{\braket{\rho|\frac{\partial}{\partial \mu}H_L|\rho}}{_{N,L}}=\lim_{L\rightarrow\infty}\frac{\partial}{\partial \mu} \frac{E^{\rho}_{N,L}}{L}\,.
\label{Eq:FH}
\ee
Commuting the thermodynamic limit with the derivative with respect to $\mu$ in the l.h.s. we have 
\be
\lim_{L\rightarrow\infty}\frac{1}{L}\tensor*[_{N,L}]{\braket{\rho|\frac{\partial}{\partial \mu}H_L|\rho}}{_{N,L}}=\left\langle\frac{\partial}{\partial \mu}\mathcal{H}(x)\right\rangle_\rho\,,\qquad \forall\,\, x \in \mathbb{R}\,,
\label{Eq:muderivative} 
\ee
where $\mathcal{H}(x)$ is the Hamiltonian density. Extra care has to be taken when performing the derivative in the r.h.s. of Eq. \fr{Eq:muderivative}. Indeed, the operator $:\!\cosh g \phi(x)\!:$, with the normal ordering prescription used in this work, is normalized in such a way that 
\be
\braket{\Omega|:\!\cosh g \phi(x)\!:|\Omega} = 1\,,
\ee 
where $\ket{\Omega}$ is the vacuum of the interacting theory. This means that \emph{$:\!\cosh g \phi(x)\!:$ has a $\mu$-dependent normalization}. In order to compute the derivative we need to explicit its $\mu$ dependence. A possible way of proceeding is to adopt for $\cosh g \phi(x)$ a different regularization procedure and hence a different normalization, which does not depend on $\mu$. A convenient choice is the ``CFT normalization", adopted, e.g., in Refs.~\cite{zamolodchikov-00}. There, the authors use the operator $[\cosh g \phi(x)]_{\rm CFT}$, normalized in such a way that 
\be
\braket{0|[\cosh g \phi(x)]_{\rm CFT}[\cosh g \phi(y)]_{\rm CFT}|0}=\frac{1}{2}|x-y|^{\frac{g^2}{2 \pi}}\,.
\ee 
Here $\ket{0}$ is the ground state of the theory for $\mu=0$. As a consequence, the operator $[\cosh g \phi(x)]_{\rm CFT}$ is independent of $\mu$.

The expectation value of $[\cosh g \phi(x)]_{\rm CFT}$ on the interacting ground state $\ket{\Omega}$, has been computed in Refs.~\cite{LZ:expval, FLZZ:expval} and reads as 
\be
\braket{\Omega|[\cosh g \phi(x)]_{\rm CFT}|\Omega}= \mathcal{G}_{g}\equiv \frac{(1+\alpha)\pi \Gamma(1+\frac{g^2}{8 \pi})}{16 \sin(\pi \alpha) \Gamma(-\frac{g^2}{8 \pi})}\left(\frac{\Gamma(\frac{1}{2}+\frac{\alpha}{2})\Gamma(1-\frac{\alpha}{2})}{4\sqrt{\pi}}\right)^{-2-\frac{g^2}{4\pi}}\left(\frac{\sin\alpha\pi}{\alpha\pi}\right)^{-\frac{g^2}{8\pi}} \mu^{-\frac{g^2}{4\pi}}\,. 
\label{Eq:expectation}
\ee 
As a consequence, we can relate the two operators as follows 
\be
:\!\cosh g \phi(x)\!:\, =\frac{1}{\mathcal{G}_{g}}[\cosh g \phi(x)]_{\rm CFT}\,.
\label{Eq:normalordered}
\ee
Using \fr{Eq:expectation} and \fr{Eq:normalordered} we find 
\be
\frac{\partial}{\partial \mu} :\!\cosh g \phi(x)\!:\, = \frac{g^2}{4 \pi \mu} :\!\cosh g \phi(x)\!:\,.
\ee
Putting all together we have 
\be
\frac{\partial}{\partial \mu}\mathcal{H}(x) = \frac{2 \mu}{g^2} \left(1+\frac{g^2}{8 \pi}\right) :\!\cosh g \phi(x)\!:\,.
\label{Eq:dHdmu}
\ee
Plugging now \fr{Eq:dHdmu} into \fr{Eq:FH} and using \fr{Eq:alphamass} gives  
\be
\braket{:\!\cosh g \phi(x)\!:}_\rho = \braket{\Phi_{1}(x)}_\rho =\lim_{L\rightarrow\infty}\left\{ 1+\frac{4 \sin(\pi \alpha)}{m}\sum_{i=1}^{N}\left\{ \cosh \theta_i + m \sinh\theta_i \frac{\partial \theta_i}{\partial m} \right\}\right\}\,.
\label{new_aux}
\ee
In the first step we used that the state $\lim_{L\rightarrow \infty} \ket{\rho}_{N,L}$ is an eigenstate of the $\mathbb{Z}_2$ symmetry \fr{Eq:Z2}.

Taking the derivative with respect to $m$ of the Bethe equations \fr{bethe_eq} one obtains an equation for the derivatives $\partial \theta_i/\partial m$:
\be
\sinh(\theta_j)+m\frac{\partial \theta_j}{\partial m}\cosh(\theta_j)=-\frac{1}{L}\sum_{k}\left(\frac{\partial \theta_j}{\partial m}-\frac{\partial \theta_k}{\partial m}\right)\varphi_{\alpha}(\theta_j-\theta_k).
\ee
Defining $f(\theta_j):=\partial \theta_j/\partial m$, it is easy to take the thermodynamic limit of the equation above, and using the Bethe equations \fr{thermo_bethe_eq} we obtain
\be
\sinh(\theta)=-2\pi f(\theta)(\rho(\theta)+\rho^h(\theta))+\int d \theta' f(\theta')\rho(\theta')\varphi_{\alpha}(\theta-\theta').
\ee
Defining now $b(\theta)=2\pi f(\theta)(\rho(\theta)+\rho^{h}(\theta))$, after straightforward steps, from Eq. \fr{new_aux} one finally obtains the final result for the expectation value of the vertex operator $\Phi_{1}(x)$
\begin{align}
\braket{\Phi_{1}(x)}_\rho &= 1+\frac{4 \sin(\pi \alpha)}{m}\int{\rm d}\theta\left\{ \rho(\theta)\cosh \theta + \frac{m}{2 \pi} \frac{1}{1+\eta(\theta)}\sinh(\theta) b(\theta)\right\}\,,\label{Eq:FHR}\\
b(\theta) &= -\sinh \theta +\int\frac{{\rm d}\theta'}{2 \pi}\left\{\frac{1}{1+\eta(\theta')}\varphi_{\alpha}(\theta-\theta') b(\theta')\right\}\,.\label{Eq:b}
\end{align}
This result is exactly the same as in Eq. \fr{Eq:main} for $k=0$. To see this we note that the ratio in the l.h.s. of Eq. \fr{Eq:main} reduces, for $k=0$, to $\braket{\Phi_{1}(x,t)}_{\rho}$ and equation \fr{Eq:p} becomes 
\begin{align}
&p_0(\theta)=e^{-\theta}+\int_{-\infty}^{\infty} \!\frac{{\rm d}\mu}{2\pi}\frac{1}{1+\eta(\mu)} \varphi_{\alpha}(\theta-\mu)p_0(\mu)\,.
\end{align}
We can now rewrite $p_0(\theta)$ as 
\be
p_0(\theta)=s(\theta)+d(\theta)\,,
\ee
where $s(\theta)$ and $d(\theta)$ satisfy
\begin{align}
&s(\theta)=\cosh\theta+\int_{-\infty}^{\infty} \!\frac{{\rm d}\mu}{2\pi}\frac{1}{1+\eta(\mu)} \varphi_{\alpha}(\theta-\mu)s(\mu)\,,\label{Eq:sum}\\
&d(\theta)=-\sinh\theta+\int_{-\infty}^{\infty} \!\frac{{\rm d}\mu}{2 \pi}\frac{1}{1+\eta(\mu)} \varphi_\alpha(\theta-\mu)d(\mu)\,.\label{Eq:diff}
\end{align}
The first equation is solved by  
\be
s(\theta)= \frac{2 \pi }{m}\left(\rho(\theta)+\rho^h(\theta)\right)\,,
\ee
while the second is identical to Eq. \fr{Eq:b}. Substituting back in Eq. \fr{Eq:main} we have 
\be
\label{eq:aux2}
{\left\langle \Phi_{1}(x)\right\rangle_{\rho}}=1+\frac{4\sin\left(\pi \alpha\right)}{m}\int_{-\infty}^{\infty} \!{\rm d}\theta	 \left\{e^{\theta} \rho(\theta)  + \frac{m}{2 \pi}\frac{e^{\theta}}{1+\eta(\theta)} b(\theta) \right\}\,.
\ee 
If $\rho(\theta)$ and hence $\eta(\theta)$ are symmetric (which is the case of interest in this work), it is trivial to see that this is the same equation as \fr{Eq:FHR}.

To prove the equivalence between the r.h.s. of Eq. \fr{Eq:FHR} and formula \fr{eq:aux2} in the case of generic, non-symmetric, distributions, we need two more steps. First note that the functions $\rho(\theta)$ and $b(\theta)$ can be expanded ``perturbatively'' in the functional parameter $r(\theta)$, given in Eq. \fr{eq:r_parameter}. The expansions read as 

\begin{align}
&b(\theta)=-\sinh\theta-\sum_{n=1}^\infty\int\ldots\int\prod_{i=1}^{n}\left(\frac{{\rm d}\theta_i}{2\pi}r(\theta_i)\right)\varphi_{\alpha}(\theta-\theta_1)\varphi_{\alpha}(\theta_1-\theta_2)\ldots \varphi_{\alpha}(\theta_{n-1}-\theta_n) \sinh(\theta_n)\,,\\
&\rho(\theta)=\frac{m}{2\pi}r(\theta) \cosh\theta +\frac{m}{2\pi} r(\theta) \sum_{n=1}^\infty\int\ldots\int\prod_{i=1}^{n}\left(\frac{{\rm d}\theta_i}{2\pi}r(\theta_i)\right)\varphi_{\alpha}(\theta-\theta_1)\varphi_{\alpha}(\theta_1-\theta_2)\ldots \varphi_{\alpha}(\theta_{n-1}-\theta_n) \cosh(\theta_n)\,.
\end{align}
Then, substituting in \fr{eq:aux2} and \fr{Eq:FHR} we construct an expansion for the r.h.s. of both equations. It is straightforward to verify that the two expansions are equivalent order by order. Hence, we showed that formula \fr{Eq:main} recovers the result obtained by the Feynman-Helmann theorem for expectation values on states described by generic distributions $\rho(\theta)$ and $\rho^h(\theta)$.

\section{Mapping to the one-dimensional Lieb-Liniger gas}\label{sec:scaling}

As mentioned in the introduction, there exists a mapping between the sinh-Gordon field theory to a non-relativistic gas of bosons with repulsive pointwise interactions, namely the Lieb-Liniger model \cite{kmt-11,kmt-09}. At an intuitive level, the latter mapping can be understood as a non-relativistic limit in which the velocity of light $c$ is sent to infinity.

It is natural to consider such a mapping within the framework of interest in this work, namely the one of quantum quenches. In particular, an intriguing question is whether mass and  interaction quenches in the sinh-Gordon model correspond, in the non-relativistic limit, to some physically relevant quenches in the Lieb-Liniger model. This question is investigated in this section. In the following we refer to \cite{kmt-09} for a complete treatment of the mapping from the sinh-Gordon field theory onto the Lieb-Liniger model, while here we only review the aspects that are relevant for our discussion.

We begin by introducing the Lieb-Liniger Hamiltonian, which reads 
\be
H=\int {\rm d}x\left\{\frac{1}{2m}\partial_x\Psi^{\dagger}(x)\partial_x\Psi(x)+\kappa\Psi^{\dagger}(x)\Psi^{\dagger}(x)\Psi(x)\Psi(x)\right\},
\label{ll_hamiltonian}
\ee
where $\kappa$ denotes the interaction strength and where  $\Psi^{\dagger}$, $\Psi$ are operators satisfying
\be
\left[\Psi(x),\Psi(y)\right]=\left[\Psi^{\dagger}(x),\Psi^{\dagger}(y)\right]=0\qquad \left[\Psi(x),\Psi^{\dagger}(y)\right]=\delta\left(x-y\right).
\ee
This model can be exactly analysed by the Bethe ansatz both in the repulsive ($\kappa\geq 0$) and attractive regime ($\kappa<0$). In the following, we focus on the repulsive regime. Note that in the attractive case an analogous mapping exists from the sine-Gordon field theory to the attractive Lieb-Liniger model \cite{ckl-14}.

In the repulsive regime, the Lieb-Liniger model does not exhibit bound states.  The excitation spectrum can then be understood in terms of a single particle species as in the sinh-Gordon field theory and the two-body scattering matrix is given by
\be
S_{\rm LL}(\lambda)=\frac{\lambda-i2m\kappa}{\lambda+i2m\kappa}.
\label{ll_s-matrix}
\ee
The TBA description is very similar to the case of the sinh-Gordon model. In particular, one defines the rapidity and hole distributions which we indicate with $\rho_{\rm LL}(\lambda)$, $\rho_{\rm LL}^{h}(\lambda)$. We also introduce $\eta_{\rm LL}(\lambda)=\rho_{\rm LL}^{h}(\lambda)/\rho_{\rm LL}(\lambda)$.

The Bethe equations in the thermodynamic limit in this case read
\be
\rho_{\rm LL}(\lambda)+\rho_{\rm LL}^h(\lambda)=\frac{1}{2\pi}+\int \frac{{\rm d}\mu}{2\pi} \varphi_{\rm LL}(\lambda-\mu)\rho_{\rm LL}(\mu),
\label{ll_bethe_eq}
\ee
where
\be
\varphi_{\rm LL}(\lambda):=\frac{{\rm d}}{{\rm d}\lambda}\left(-i\ln S_{\rm LL}(\lambda)\right)=\frac{4m\kappa}{\lambda^2+4m^2\kappa^2}.
\label{Eq:varphi1LL}
\ee
Note that the form of Eq. \fr{ll_bethe_eq} is analogous to that of Eq. \fr{thermo_bethe_eq}.  Going further, we now perform the following rescaling of the sinh-Gordon root densities
\bea
\tilde{\rho}(\lambda)&=&\frac{1}{mc\cosh\theta(\lambda)}\rho(\theta(\lambda)),\label{tilde_function_I}\\\qquad \tilde{\rho}^h(\lambda)&=&\frac{1}{mc\cosh\theta(\lambda)}\rho^h(\theta(\lambda)),
\label{tilde_function_II}
\eea
where $\theta(\lambda)$ is defined by
\be
\lambda=m c\sinh\theta(\lambda).
\ee
It is immediate to see that the distributions $\tilde{\rho}$, $\tilde{\rho}^h$ satisfy
\be
\tilde{\rho}(\lambda)+\tilde{\rho}^h(\lambda)=\frac{1}{2\pi}+\int_{-\infty}^{\infty}\frac{{\rm d}\mu}{2\pi}\widetilde{\varphi}_{\alpha}(\lambda,\mu)\tilde{\rho}(\mu),
\label{thermo_bethe_eqLL}
\ee
where
\be
\tilde{\varphi}_{\alpha}(\lambda,\mu)=\frac{1}{mc\cosh(\theta(\lambda))}\frac{2\cosh(\theta(\lambda)-\theta(\mu))\sin(\alpha\pi)}{\sinh^2(\theta(\lambda)-\theta(\mu))+\sin^2(\alpha\pi)}.
\label{Eq:varphiLL}
\ee
Eq. \fr{thermo_bethe_eqLL} has now the same form of Eq. \fr{ll_bethe_eq}, where the only difference is given by the kernel \fr{Eq:varphiLL}. The Bethe equations in the Lieb-Liniger model are then exactly recovered by the following double scaling limit:
\bea
g&\to& 0,\notag\\
c&\to& \infty,\label{scaling_limit_II}\\
gc&=&4\sqrt{\kappa}={\rm fixed.}\notag
\eea
where we remind that $c$ is the speed of light, and where $g$ is the parameter appearing in the Hamiltonian \fr{hamiltonian}. Indeed, under the above double scaling limit one can immediately see that $\tilde{\varphi}_{\alpha}(\lambda,\mu)\to\varphi_{\rm LL}(\lambda-\mu)$.

As it is thoroughly discussed in Refs.~\cite{kmt-11, kmt-09}, the mapping encoded in Eqs. \fr{scaling_limit_II} goes beyond the TBA equations, and it also involves a direct correspondence between the scattering matrices and the field operators in the sinh-Gordon field theory and the Lieb-Liniger model. Since in this section we restrict to the analysis of relations between rapidity distributions, we refer to \cite{kmt-11,kmt-09} for more details on these aspects of the mapping defined above. 

As we already pointed out, it is now natural to apply the double scaling limit in \fr{scaling_limit_II} to the integral equation defining the long-times stationary state after the mass and interaction quench discussed in section \ref{sec:initial}. If we take the double limit in \fr{scaling_limit_II} of Eq. \fr{e_saddle_point}, and bring the limit inside the integral in the r.h.s., we straightforwardly obtain 
\be
\ln \eta_{\rm LL}(\lambda)=-\ln \left(\frac{m_0-m}{m_0+m}\right)^2-\ln\left[\frac{\lambda^2}{\lambda^2+m^2\kappa^2}\right]-\int_{-\infty}^{\infty}\frac{{\rm d}\mu}{2\pi}\varphi_{\rm LL}(\lambda-\mu)\ln\left(1+\frac{1}{\eta_{\rm LL}(\mu)}\right)\,.
\label{scaled_eq}
\ee
Interestingly, this equation is very similar to the one defining the stationary state in the quench from a BEC state in the Lieb-Liniger model found in Ref.~\cite{dwbc-14}. In particular, we see that the only difference lies in the driving term. However, the latter is the logarithm of a function which is not vanishing for large $\lambda$, namely
\be
\lim_{\lambda\to\infty}\frac{\lambda^2}{\lambda^2+m^2\kappa^2}=1.
\label{eq:aux_2}
\ee
Note that this is different to the case of Ref. \cite{dwbc-14} and in fact Eq. \fr{eq:aux_2} has serious consequences. Indeed, it can be seen that because of \fr{eq:aux_2} the inverse of the solution $\eta^{-1}_{\rm LL}(\lambda)$ is not vanishing for $\lambda\to\infty$. In turn, this implies
\be
\lim_{\lambda\to\infty}\rho_{\rm LL}(\lambda)\neq 0,
\ee
namely the solution of Eqs. \fr{ll_bethe_eq} and \fr{eq:aux_2} corresponds to an infinite density of particles. Heuristically, this can be interpreted as the signal that in the non-relativistic limit $c\to\infty$ a mass quench pumps an infinite energy into the system and an infinite number of particles is produced.

\begin{figure}
\includegraphics[scale=0.85]{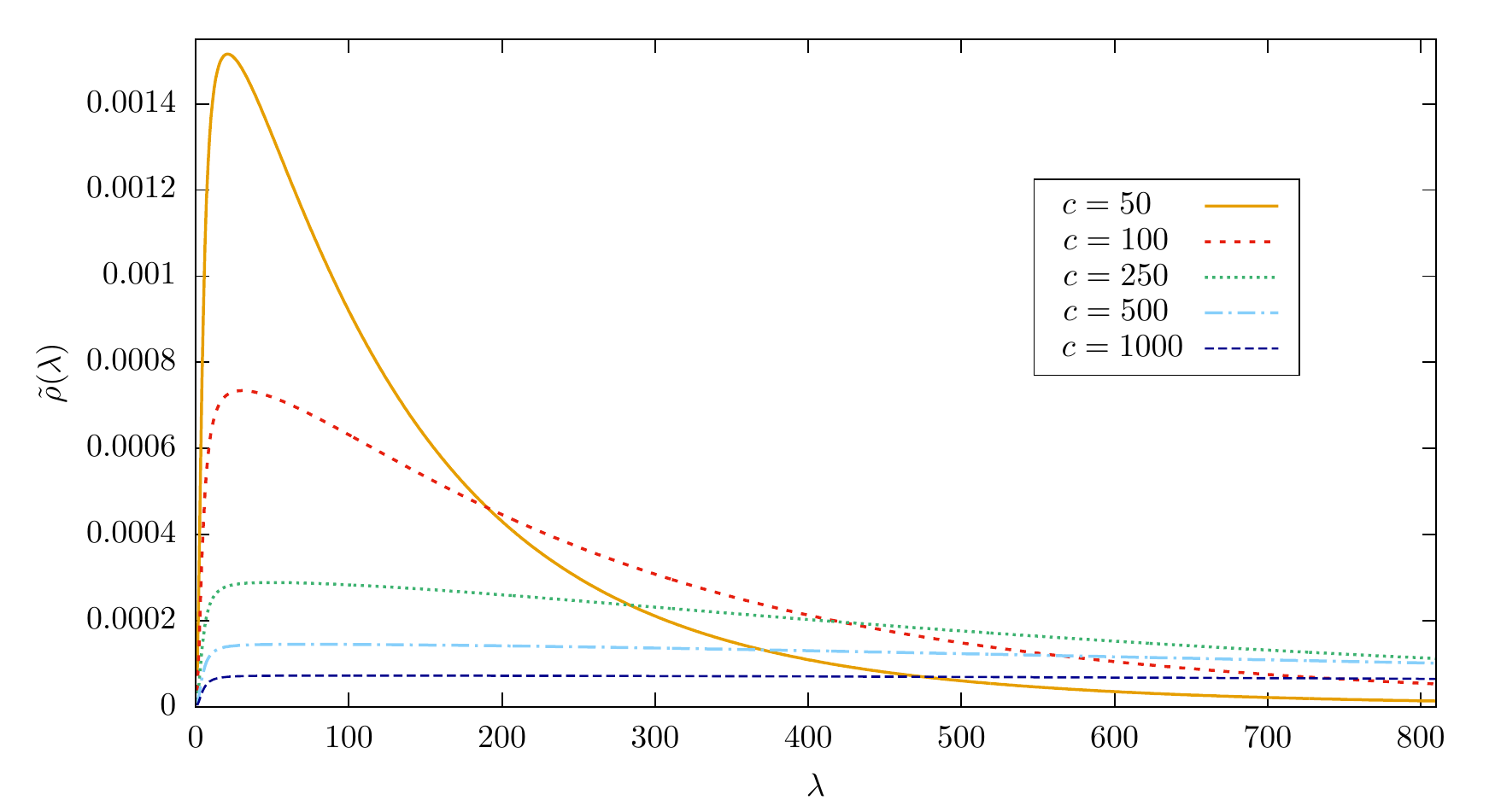}
\caption{Scaled distribution $\tilde{\rho}(\lambda)$ [cf. Eq. \fr{tilde_function_I}] corresponding to the stationary state reached after the quench described in section \ref{sec:initial}. The ratio between initial and final masses in the sinh-Gordon model is chosen to be $m/m_0=0.1$. Here we chose $\kappa=1$, while the density is kept fixed to the value $D=0.5$ by approprietly tuning the Lagrange multiplier in Eq. \fr{lagrange_mul}. We see that, as $c$ increases, the distribution flattens and its width increases, the value of $\tilde{\rho}(\lambda)$ eventually approaching zero at any finite $\lambda$.}\label{fig:scaling}
\end{figure}

In order to fix the density of particles, a different procedure is required, namely one has to introduce a Lagrange multiplier (dependent on $c$) in the saddle-point equation \fr{e_saddle_point}
\be
\ln \eta_{\rm sp}(\theta)=-h(c)-\ln|K(\theta)|^2-\int_{-\infty}^{\infty}\frac{{\rm d}\theta'}{2\pi}\varphi_{\alpha}(\theta-\theta')\ln\left(1+\frac{1}{\eta_{\rm sp}(\theta')}\right).
\label{lagrange_mul}
\ee
The Lagrange multiplier $h(c)$ is chosen in such a way to keep the density constant under the limit \fr{scaling_limit_II}, i.e.
\be
\int_{-\infty}^{\infty}\tilde{\rho}(\lambda){\rm d}\lambda=D\, ,\qquad {\rm (independent\ of\ }c{\rm )}.
\ee
Note that in the double scaling limit, $h(c)$ will be divergent as it was also the case for the Lagrange multiplier in the equilibrium setting studied in Refs.~\cite{kmt-11,kmt-09}.

One can now solve Eqs. \fr{lagrange_mul} and consequently \fr{thermo_bethe_eq} for increasing values of $c$, by fixing the density $D$ and
\be
g=\frac{4\sqrt{\kappa}}{c},
\ee
according to Eq~\fr{scaling_limit_II}. The resulting rapidity distributions $\rho(\theta)$ can be plugged in \fr{tilde_function_I} to obtain the corresponding distribution $\tilde{\rho}(\lambda)$. The result of this procedure is shown in Fig. \ref{fig:scaling}. It can be clearly seen that, as $c$ increases, the distribution $\tilde{\rho}(\lambda)$ flattens and its width increases. Heuristically, this corresponds to the fact that the particles in the system acquire larger kinetic energy as the value of $c$ is increased. 

In fact, a natural energy scale for the energy per particle produced during the quench is $\Delta m c^{2}$ which goes to infinity as $c$ increases (here, $\Delta m$ is the difference between the initial and final masses $m_0$, $m$). In the limit $c\to \infty$ each particle acquires infinite kinetic energy, meaning that the procedure described above does not have a well defined limit, as it would yield $\rho_{\rm LL}(\lambda)=0$, for any finite $\lambda$.

From this discussion, one might expect that a well defined non-relativistic limit is obtained considering an infinitesimal mass quench, namely $m_0=m+\Delta(c)$, with $\Delta(c)$ some quantity which is vanishing for $c\to \infty$. However, we explicitly verified that this is not the case for the initial state \fr{initial} defined by the function $K(\theta)$ in Eq. \fr{k_function}.

In summary, we have seen that the specific mass quench considered in this work does not have a well defined non-relativistic limit, thus it does not yield a well defined quench situation in the Lieb-Liniger model. This is to be ascribed to the intrinsic relativistic nature of a mass quench in the sinh-Gordon field theory which naturally implies particle production. It is nevertheless reasonable to expect that different quench protocols in the sinh-Gordon field theory could yield well defined non-relativistic limits. An example might be provided by interaction quenches where the initial state is chosen to be the ground state of the Hamiltonian \fr{hamiltonian} with a non-zero chemical potential. The main obstacle in studying these situations is, once again, the difficulty to find the expansion of the initial state in the basis of the post-quench Hamiltonian.

\section{Conclusions}\label{sec:conclusions}

In this work we have considered quantum quenches in the sinh-Gordon model, focussing on a particular class of initial states. We used the quench action approach to fully characterize the steady state which describes the late-time behaviour of local observables, in terms of the corresponding rapidity and hole distributions. We have verified that the predictions for the local observables match the ones obtained using the GGE involving the conserved rapidity occupation numbers, in agreement with the results of Refs.~\cite{fioretto-10, mussardo, pozsgay-11}. We have considered the computation of one-point correlation functions on the post-quench stationary state, for which we have proposed a formula based on the recent finite temperature results of Refs.~\cite{ns-13, negro-14}. Finally, we have considered the mapping of the sinh-Gordon field theory to the Lieb-Liniger model, previously investigated only within equilibrium settings, and we have studied its application to the quench problem discussed in this manuscript.

Our work provides a non-trivial application of the QAM, to the study of interacting quantum quenches in integrable field theories. Note that the simple structure of the excitations of the sinh-Gordon model has allowed here to exactly characterize the post-quench stationary state (with the initial condition given by Eq. \fr{initial}). This is to be compared with the analogous quench in the sine-Gordon field theory investigated in Ref. \cite{bertini}, where the richer structure of the model only allows for an analysis at the first order in the density of excitations.

As we have shown, the study of quantum quenches in the sinh-Gordon field theory has, potentially, ramifications also for the Lieb-Liniger gas. Indeed, from our analysis it seems reasonable to expect that particular quenches in the sinh-Gordon field theory could be identified, in the non-relativistic limit, with physically meaningful quench settings in the Lieb-Liniger model. This scenario might provide computational advantages in the determination of the post-quench stationary correlation functions in the Lieb-Liniger gas, as it was the case for the equilibrium situations studied in Ref. \cite{kmt-11,kci-11,kmt-09}.

In this respect, it would also be interesting to investigate the possibility of exploiting the formula proposed in this work to obtain explicit results for one-point correlation functions of arbitrary excited states in the Lieb-Liniger gas. Indeed, up to now, in the latter model explicit efficient formulae for the $K$-body one-point correlation functions are known only for $K\leq 4$ \cite{kci-11, pozsgay2-11}. This issue will be addressed in future studies.

A natural question arising from our work concerns the computation of the whole time evolution of local observables after the quench. Even though this is in principle within the reach of the quench action approach, it is in general still extremely hard to derive explicit results, which have been obtained only in a few special cases \cite{bertini, dc-14,pdc-15, vwed-15}. In  Ref.~\cite{pdc-15}, within the framework of the QAM and exploiting efficient formulae for the form factors of local operators as obtained in Refs. \cite{pozsgay2-11, pc-15}, a numerical evaluation was performed to exactly compute the time evolution of a particular one-point function in the Lieb-Liniger gas. In our case, despite the similarities between the Bethe ansatz description of the latter model and of the sinh-Gordon field theory, a more sophisticated treatment might be needed, because of the presence of singularities in the form factors of vertex operators \cite{cef-11, se-12, bertini}. A systematic analysis of this point goes however beyond the scope of the present work and will be investigated in the future.

\section{Acknowledgments}
We thank Giuseppe Mussardo and Spyros Sotiriadis for useful discussions surrounding this work. B.B. and P.C. acknowledge the financial support by the ERC under Starting Grant 279391 EDEQS.
\appendix

\section{Connected form factors}\label{App:CFF}

In this appendix we report explicitly the connected diagonal $n$-particle form-factors of the vertex operators $\Phi_k(x)$ defined in Eq. \fr{Eq:vertex} for $n=1,2,3$. They are straightforwardly derived by the explicit expressions of the form-factors of $\Phi_k(x)$, which have been found in Ref. \cite{KM:FF}. Indeed, one simply needs to follow the prescription sketched in section \ref{sec:LM} and perform the limit \fr{Eq:limit}. The result reads 
\begin{align}
&\braket{\theta_1|\Phi_k(0)|\theta_1}_c =4\sin(\pi \alpha)[k]^2\\
&\braket{\theta_2,\theta_1|\Phi_k(0)|\theta_1,\theta_2}_c =\frac{16 [k]^2 \sin^2(\pi \alpha)}{\sin^2(\pi \alpha)+\sinh^2(\theta_1-\theta_2)}\left(\cosh^2(\theta_1-\theta_2)[k]^2-[k-1][k+1]\right)\\
&\braket{\theta_3,\theta_2,\theta_1|\Phi_k(0)|\theta_1,\theta_2,\theta_3}_c = 2 [k] \prod_{i<j =1}^{3}\left\{\frac{\sin(\pi \alpha)}{\sin^2(\pi \alpha)+\sinh^2(\theta_i-\theta_j)}\right\}{\mathcal F}_3(\theta_1,\theta_2,\theta_3)\,.
\end{align} 
Here 
\be
[k]\equiv\frac{\sin(\pi \alpha k)}{\sin(\pi \alpha)},
\ee
and 
\begin{align}
{\mathcal F}_3(\theta_1,\theta_2,\theta_3)\equiv& \bigl\{11\cosh(2 \theta_{12}) +\cosh(4 \theta_{12}) +2 \cosh(2 \theta_{321}) + 11 \cosh(2 \theta_{13}) \notag\\
&+ \cosh(4 \theta_{13})
+ 11 \cosh(2 \theta_{23}) + \cosh(4 \theta_{23}) +2 \cosh(2 \theta_{123}) +2 \cosh(2 \theta_{213}) + 12\bigr\}[k]^5\notag\\
&+ 2 [k-1][k][k+1]\bigl\{(9 + 7 \cosh(2 \theta_{12}) + 7 \cosh(2 \theta_{13}) + 
        7 \cosh(2 \theta_{23})) [k-1][k+1] - 3 [k-2][k+2]\bigr\} \notag\\
&+ [k]^3 \bigl\{-2 (15 + 14 \cosh(2 \theta_{12}) + 2 \cosh(2 \theta_{312}) + 14 \cosh(2 \theta_{13}) + 14 \cosh(2 \theta_{23})  + 
        2 \cosh(2 \theta_{231})\notag\\
&  + 2 \cosh(2 \theta_{123}))[k-1][k+1] -(3 + \cosh(2 \theta_{12}) + \cosh(2 \theta_{13})+\cosh(2 \theta_{23})) [k-2][k+2]\bigr\}\notag\\
&+2 \{3 + 2 \cosh(2 \theta_{12}) +2\cosh(2 \theta_{13})+2 \cosh(2 \theta_{23})\}[k]^2 \{[ k-2][k+1]^2 + [k-1]^2 [k+2]\}\notag\\
&-\{3 + \cosh(2 \theta_{12})+ \cosh(2 \theta_{13}) +\cosh(2 \theta_{23})\} [k-1][k+1] \{[k-2][k+1]^2 + [k-1]^2 [k+2]\}\,,
\end{align}
where we introduced $\theta_{ij}\equiv \theta_{i}-\theta_{j}$ and $\theta_{ijk}\equiv 2\theta_{i}-\theta_{j}-\theta_{k}$.

\end{document}